# Dust productivity and impact collision of the asteroid (596) Scheila

L. Neslusan [a,*], O. Ivanova [b], M. Husarik [a], J. Svoren [a], Z. Seman Krisandova [a]

[a] Astronomical Institute, Slovak Academy of Sciences, 05960 Tatranská Lomnica, Slovakia
[b] Main Astronomical Observatory of NAS of Ukraine, Akademika Zabolotnoho 27, 03680 Kyiv, Ukraine

## ABSTRACT

Photometric observations of asteroid (596) Scheila were obtained during and after its 2010 outburst. The estimated radius of the body (spherical approximation of the asteroidal body) was $51.2 \pm 3.0$ km and $50.6 \pm 3.0$ km for different methods. The ejected dust mass from the asteroid ranged from $2.5 \times 10^7$ to $3.4 \times 10^7$ kg for different methods. An impact mechanism for triggering Scheila's activity is discussed. A few days before the impact, Scheila passed through the corridors of two potential cometary streams.

## Introduction

The main-belt asteroid (596) Scheila (1906 UA) was discovered by A. Kopff in Heidelberg in 1906. It is classified as the Tholen PCD class or SMASSII T class asteroid (Bus and Binzel, 2002). Licandro et al. (2011) showed that its surface was homogeneous and corresponded to a dark primitive D-type asteroid. It is a middle-sized body with the diameter of 113 km and visual geometric albedo of 0.038 (Tedesco and Desert, 2002). The asteroid moves in the orbit with semi-major axis 2.938 AU, eccentricity 0.165, and inclination 14.7°. Its orbit is situated in the outer asteroid belt (Fig. 1), where several main-belt comets (Hsieh and Jewitt, 2006) have been found.

Similarly, Scheila is a typical asteroid according to the value of its Tisserand parameter with respect to Jupiter, which equals $T_J = 3.21$. The observations taken with the 0.68-m Catalina Schmidt telescope on December 11.44–11.47 (UT), 2010, showed a cometary activity (Larson et al., 2010). The archival Catalina observations showed that the activity was triggered before December 3, when Scheila appeared as a slightly diffuse object ($V = 13.2$ mag). With a main-belt orbit ($T_J > 3$) and a comet-like morphology, Scheila obeys the definition of a main-belt comet or active asteroid (Hsieh and Jewitt, 2006). At the time of outburst, Scheila was the seventh known example of this specific class of the objects in the Solar System. The currently known main-belt comets (MBCs, hereinafter) are 133P/Elst-Pizarro, 176P/LINEAR, 238P/Read, 259P/Garradd, P/2010 A2 (LINEAR), 324P/La Sagra, 2006 VW139, P/2012 F5 (Gibbs), P/2012 T1 (PANSTARRS), 311P/PANSTARRS, and P/2013 R3 (Catalina-PANSTARRS).

Active asteroids are now recognized as a new class of objects in the Solar System. These objects are remarkable since they have both the orbital characteristics of asteroids and the physical characteristics of comets. It means that they look like comets because they show comae and tails, but they have the whole orbits inside the Jupiter's orbit. In the literature, sublimation, impact, electrostatics and rotational bursting, thermal effects, and radiation pressure sweeping have been proposed as the mechanisms of their activity (Hsieh and Jewitt, 2006; Jewitt, 2012; Jewitt et al., 2016).

Because of their activity, the MBCs seem to be related to the objects in the well-established comet reservoirs, Oort cloud and Kuiper belt. However, there is no dynamical pathway between two distant reservoirs and the MBCs in the Solar System.

In the period of activity of Scheila, its spectra were obtained and emissions were searched for. Bodewits et al. (2011) showed that no gases were detected by Swift UV-optical observations, suggesting that the outburst was not triggered by the ice sublimation. Other physical evidence suggests (Yang and Hsieh, 2011; Jewitt et al., 2011; Ishiguro et al., 2011) that Scheila likely collided, recently, with another, smaller asteroid, and that its activity is unlikely produced by sublimation.

Since Scheila is the largest known active asteroid, or MBC, it was a suitable candidate also for observations performed at the Skalnaté Pleso Observatory during the period of its activity. In 2010 and 2011, Husárik (2012) performed a relative photometry to compare and determine whether and, if so, how much the rotational period was changed by an outburst with respect to the data published by Warner in 2006. The comparison shows very similar

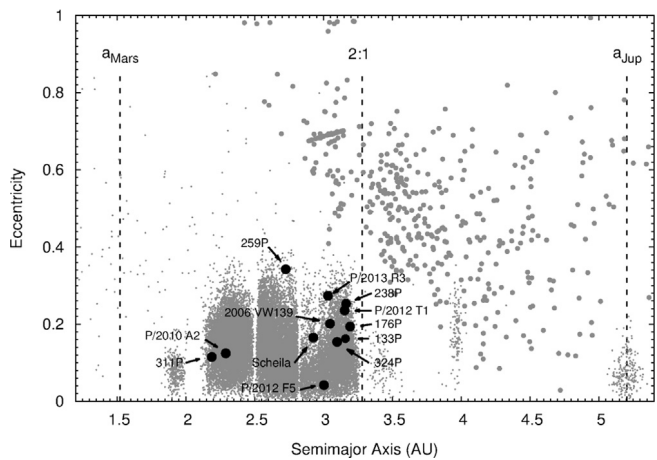

**Fig. 1.** The distribution of the known main-belt comets (big black circles), asteroids (grey dots), and comets (bigger grey dots) in the semi-major axis vs. eccentricity plane. The dashed lines show the semi-major axes of the orbits of Mars, Jupiter, and the location of the 2:1 mean-motion resonance with Jupiter.

lightcurves and, hence, no significant change in the rotational period.

In our article, we analyze the photometric data of the Scheila in the period of its activity and after. We estimate the dust productivity in the outburst phase and radius of the asteroid after this period, when object was not active.

Furthermore, we expanded our work to find possible sources of the impact, which triggered the Scheila's sudden activity. In principle, there are three source regions of the impactor: (i) an individually moving small asteroid, fragment of comet, or sporadic meteoroid, (ii) a meteoroid from a cometary or asteroidal stream, and (iii) a component of a known binary asteroid. In the first case, we cannot identify the previously unknown impactor and, thus, say anything more about the circumstances of the collision.

However, another research is possible, if the impactor is a member of a meteoroid stream or satellite of known minor body still existing in the Solar System. Because of this possibility, we look for an appropriate, cometary and asteroidal, potential parent body of meteoroid stream that could be crossed by Scheila in time of the outburst. As well, we calculate the approaches of all small bodies, with well-known orbits, to Scheila in the most probable time of the collision.

We also predict the dates of future passages of Scheila near the former collisional point of its orbit. If the impactor was a meteoroid from a stream, a certain, at least indicative, activity could again appear at the crossing of the stream corridor.

## Observation and reduction

The observations of asteroid Scheila were obtained between December 15, 2010, and May 22, 2011, with the 0.61-m f/4.3 reflector at the Skalnaté Pleso Observatory. The photometric data of Scheila were obtained through the V and R broadband filters. The CCD detector SBIG ST-10XME with $3 \times 3$ binning and resolution of 1.6 arcsec/px was used. The reduction of the raw data, which included bias subtraction, dark and flat field correction, and removal of cosmic ray tracks, was made. The morning sky was exposed to provide flat field corrections for the non-uniform sensitivity of the CCD chip. We applied the standard calibration frames with IRAF tools. The observational circumstance for the asteroid in each observational night are listed in Table 1 and Fig. 2.

**Table 1**
List of our observations of (596) Scheila. $r$ and $\Delta$ are the heliocentric and geocentric distances, respectively, $t_{exp.}$ is the duration of exposure, and $n$ is the number of exposures in a given sequence.

| Observation time (UT) | $r$ (AU) | $\Delta$ (AU) | Phase angle (deg) | $t_{exp.}$ (s) | $n$ | Filters |
|---|---|---|---|---|---|---|
| 2010 12 15.0 | 3.104 | 2.496 | 16.0 | 180 | 72 | R |
| 2010 12 16.1 | 3.102 | 2.482 | 15.8 | 100 | 40 | R |
| 2010 12 17.1 | 3.101 | 2.469 | 15.7 | 100 | 63 | R |
| 2010 12 27.9 | 3.085 | 2.338 | 13.7 | 180 | 53 | R |
| 2011 01 04.0 | 3.074 | 2.263 | 12.1 | 100 | 92 | R |
| 2011 01 06.0 | 3.071 | 2.244 | 11.6 | 180 | 93 | R |
| 2011 01 09.9 | 3.065 | 2.209 | 10.7 | 180 | 53 | R |
| 2011 01 26.8 | 3.040 | 2.105 | 7.0 | 180 | 5 | V |
| 2011 01 28.1 | 3.037 | 2.097 | 6.7 | 120 | 5 | V |
| 2011 02 01.1 | 3.029 | 2.082 | 6.2 | 120 | 5 | V |
| 2011 03 08.9 | 2.982 | 2.133 | 11.6 | 100 | 5 | V |
| 2011 03 22.7 | 2.952 | 2.272 | 16.1 | 100 | 5 | V |
| 2011 03 23.7 | 2.943 | 2.318 | 17.1 | 100 | 5 | V |
| 2011 04 02.9 | 2.933 | 2.377 | 18.0 | 100 | 5 | V |
| 2011 04 22.8 | 2.900 | 2.594 | 20.0 | 100 | 5 | V |
| 2011 05 06.8 | 2.877 | 2.753 | 20.4 | 100 | 5 | V |
| 2011 05 08.8 | 2.873 | 2.776 | 20.4 | 120 | 5 | V |
| 2011 05 22.8 | 2.850 | 2.932 | 20.0 | 100 | 5 | V |

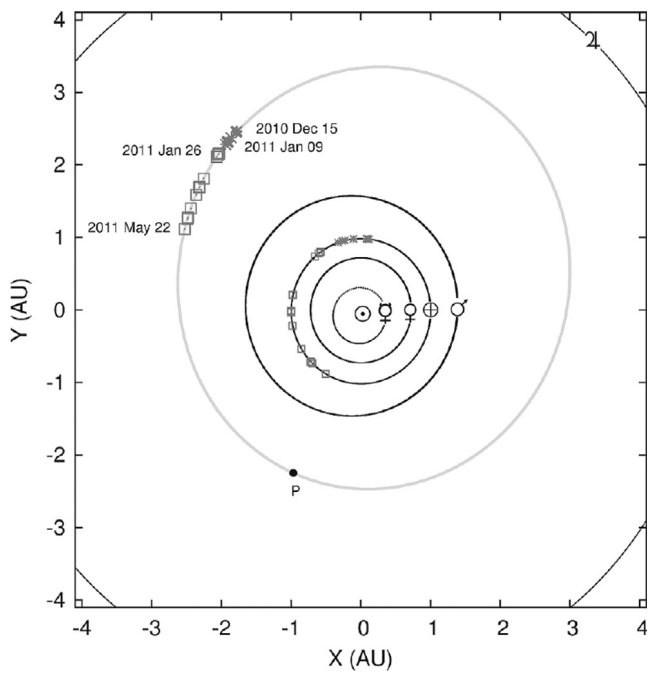

**Fig. 2.** The Scheila's orbit projected in the ecliptic plane. The asterisks correspond to our observational dates from December 15, 2010 to January 9, 2011 and empty squares correspond to the dates from January 26, 2011 to May 22, 2011. $P$ indicates the position of perihelion of the Scheila's orbit. The orbits of the terrestrial planets and Jupiter are also plotted for a comparison.

## Morphology of (596) Scheila during outburst

An impact has been proposed as the cause of the activity of (596) Scheila, in the literature. The modeling of the outburst of this asteroid implied an estimate of the impact occurrence on November 27, with an estimated accuracy of $\pm 3$ days (Moreno et al., 2011). Assuming the same density for the impactor and asteroid ($\rho = 1500$ kg m$^{-3}$), the authors estimated the projectile mean radius in the range of 30–90 m. Ishiguro et al. (2011), analyzing the changes of shape of the light curve, estimated the mass of ejecta to be $(1.5–4.9) \times 10^8$ kg, suggesting that equivalent mass of 500–800 m crater was excavated at the event. The value of impactor size (20–50 m) obtained by the authors is consistent with the impactor size derived by Jewitt et al. (2011) and Bodewits et al. (2011). We observed Scheila during its activity from December 15, 2010 to January 9, 2011. Our observations of the object showed the outburst with a structure in dust environment. To recognize the weakly contrasted structures within the images of the dust coma, we used the special software Astroart.[1] It can be applied for a number of digital filters: the Larson–Sekanina filter (Larson and Sekanina, 1984), the unsharp masking, and the Gaussian blur. The detailed information on the digital filters can be found in the paper by Ivanova et al. (2012). Such the technique of recognizing the structures in the coma was successfully used by Manzini et al. (2007) for comet C/2002 C1 (Ikeya–Zhang), Korsun et al. (2008, 2010) for distant comets, and by Ivanova et al. (2012) for comet 29P/Schwassmann–Wachmann 1. In Fig. 3 we present the results of processing of images.

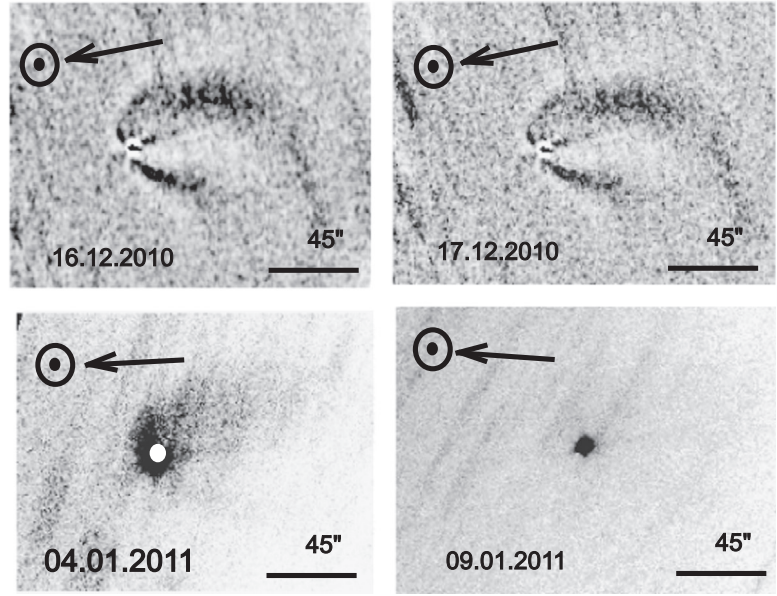

**Fig. 3.** R band images of (596) Scheila taken in six nights and processed with digital filters. The structure of outburst is changed during the time of observation. After January 9, 2011, we could not detect the dust environment. We observed Scheila as a stellar object.

**Photometry of the asteroid**

We used the observations of Scheila obtained through the R broadband filter to calculate the magnitude in the period of activity and through the V broadband filter after this period. A radius of the nucleus and the dust production were also obtained.

In the period of the outburst, Scheila was more similar to a comet than an asteroid, therefore a cometary description is more relevant. The cometary magnitude is determined by

$$m_a(\lambda) = -2.5 \log_{10}\left[\frac{I_a(\lambda)}{I_s(\lambda)}\right] + m_{st} - 2.5 \log_{10}[P(\lambda)]\Delta M, \quad (1)$$

where $m_{st}$ is the magnitude of the standard star, $I_s$ and $I_a$ are the measured fluxes of the star and the asteroid in counts, respectively, $P$ is wavelength dependent sky transparency, $\Delta M$ is the difference between the comet and star airmasses. As we used the field stars for calibration, the sky transparency is not considered.

---
[1] http://www.msb-astroart.com/.

We estimate the absolute magnitude of the object, $m_a(1,1,0)$, i.e. the magnitude being corrected to $r = \Delta = 1$ AU and phase angle $\alpha = 0°$, by using the H, G1, G2 model by Penttilä et al. (2016) (see also Muinonen et al., 2010). According to this model, the value of the magnitude, at phase angle $\alpha$ between the Sun and the observer as seen from the object, reduced to unit distance is

$$m_a(1,1,\alpha) = m_a(1,1,0) - 2.5 \log_{10}[G_1\Phi_1(\alpha) + G_2\Phi_2(\alpha) \\ + (1 - G_1 - G_2)\Phi_3(\alpha)]. \quad (2)$$

At the same time, the reduced magnitude $m_a(1,1,\alpha)$ is related to the observed magnitude of object, which is in the heliocentric and geocentric distances $r$ and $\Delta$ (in AU), respectively, and seen under the space angle $\alpha$, as

$$m_a(1,1,\alpha) = m_a(r,\Delta,\alpha) - 5 \log_{10}(r\Delta), \quad (3)$$

The basic functions $\Phi_1$, $\Phi_2$, and $\Phi_3$ in relation (2) were given in the tabulated form by Penttilä et al. (2016) (or, the parameters necessary to evaluate relation (2) can be gained using the calculator provided by these authors). The results for Scheila are given in Table 2.

*Radius of the asteroid*

We estimated the radius of the asteroid, using our observations obtained through V broadband filter in the post-activity period (March 8.9, 2011), when the object had not longer any detectable dust "coma", and looked like a typical asteroid. On March 2, 2011, Ishiguro et al. (2011) detected a faint straight tail connected to Scheila. This structure was not, however, seen in our images.

Specifically, we calculate the radius by using two formulas. To estimate the effective radius considering the cometary appearance of our object, we use relation by Meech et al. (2009):

$$p_v R_a^2 = 2.24 \times 10^{22}\, 10^{0.4[m_\odot - m_a(1,1,0)]}. \quad (4)$$

Here, $p_v = 0.038 \pm 0.040$ is the albedo of asteroid (Tedesco and Desert, 2002); $m_\odot$ is the magnitude of the Sun (the latter being $\sim -27.094$ for R band and $\sim -26.74$ for V band, (Holmberg et al., 2006).

We also estimated the diameter using the formula for an asteroid-like object, which describes the relationship between the flux and size and albedo of asteroid. This relationship is given by

**Table 2**
The absolute magnitudes of asteroid (596) Scheila determined from its photometry. The used symbols are explained in the text.

| Observation time (UT) | $r$ (AU) | $\Delta$ (AU) | Phase angle (deg) | $m_a$ | | $m_a(1,1,0)$ | | Filters |
|---|---|---|---|---|---|---|---|---|
| | | | | 11.12″ | 66.72″ | 11.12″ | 66.72″ | |
| 2010 12 15.0 | 3.104 | 2.496 | 16.0 | 13.74 ± 0.01 | 13.35 ± 0.02 | 8.72 ± 0.01 | 8.33 ± 0.02 | R |
| 2010 12 16.1 | 3.102 | 2.482 | 15.8 | 13.68 ± 0.01 | 13.32 ± 0.01 | 8.67 ± 0.01 | 8.31 ± 0.01 | R |
| 2010 12 17.1 | 3.101 | 2.469 | 15.7 | 13.71 ± 0.01 | 13.39 ± 0.01 | 8.71 ± 0.01 | 8.39 ± 0.01 | R |
| 2010 12 27.9 | 3.085 | 2.338 | 13.7 | 13.34 ± 0.02 | 13.06 ± 0.07 | 8.44 ± 0.01 | 8.16 ± 0.07 | R |
| 2011 01 04.0 | 3.074 | 2.263 | 12.1 | 13.27 ± 0.01 | 13.01 ± 0.02 | 8.41 ± 0.01 | 8.16 ± 0.02 | R |
| 2011 01 06.0 | 3.071 | 2.244 | 11.6 | 13.25 ± 0.01 | 13.08 ± 0.01 | 8.41 ± 0.01 | 8.25 ± 0.01 | R |
| 2011 01 09.9 | 3.065 | 2.209 | 10.7 | 13.28 ± 0.01 | 13.15 ± 0.02 | 8.46 ± 0.01 | 8.33 ± 0.02 | R |
| 2011 01 26.8 | 3.040 | 2.105 | 7.0 | 13.37 ± 0.01 | – | 8.57 ± 0.01 | – | V |
| 2011 01 28.1 | 3.037 | 2.097 | 6.7 | 13.33 ± 0.01 | – | 8.54 ± 0.01 | – | V |
| 2011 02 01.1 | 3.029 | 2.082 | 6.2 | 13.27 ± 0.01 | – | 8.48 ± 0.01 | – | V |
| 2011 03 08.9 | 2.982 | 2.133 | 11.6 | 13.80 ± 0.01 | – | 9.14 ± 0.01 | – | V |
| 2011 03 22.7 | 2.952 | 2.272 | 16.1 | 13.99 ± 0.01 | – | 9.29 ± 0.01 | – | V |
| 2011 03 23.7 | 2.943 | 2.318 | 17.1 | 13.88 ± 0.01 | – | 9.16 ± 0.01 | – | V |
| 2011 04 02.9 | 2.933 | 2.377 | 18.0 | 14.01 ± 0.01 | – | 9.25 ± 0.01 | – | V |
| 2011 04 22.8 | 2.900 | 2.594 | 20.0 | 14.34 ± 0.01 | – | 9.45 ± 0.01 | – | V |
| 2011 05 06.8 | 2.877 | 2.753 | 20.4 | 14.36 ± 0.01 | – | 9.36 ± 0.01 | – | V |
| 2011 05 08.8 | 2.873 | 2.776 | 20.4 | 14.43 ± 0.01 | – | 9.40 ± 0.01 | – | V |
| 2011 05 22.8 | 2.850 | 2.932 | 20.0 | 14.31 ± 0.01 | – | 9.19 ± 0.01 | – | V |

**Table 3**
The measured magnitude, effective scattering cross-section of coma, and mass of ejected dust from our R-band photometry of asteroid (596) Scheila. The result of other authors is also given for a comparison. The symbols are explained in the text.

| Observation time (UT) | $r$ (AU) | $\Delta$ (AU) | Phase angle (deg) | $\Delta m$ | $10^4 C_C$ (km$^2$) | $10^7 M_d$ (kg) | $10^7 M_d$ (kg) | | Filter | Source |
|---|---|---|---|---|---|---|---|---|---|---|
| | | | | | | | $n=3.0$ | $n=3.5$ | | |
| 2010 12 15.0 | 3.104 | 2.496 | 16.0 | 1.13 | 1.69 | 3.29 | 2.51 | 2.48 | R | This work |
| 2010 12 16.1 | 3.102 | 2.482 | 15.8 | 1.09 | 1.60 | 3.21 | 2.49 | 2.44 | R | This work |
| 2010 12 17.1 | 3.101 | 2.469 | 15.7 | 1.11 | 1.64 | 3.29 | 2.50 | 2.45 | R | This work |
| 2010 12 27.9 | 3.085 | 2.338 | 13.7 | 1.07 | 1.55 | 3.11 | 2.52 | 2.51 | R | This work |
| 2011 01 04.0 | 3.074 | 2.263 | 12.1 | 1.01 | 1.42 | 2.84 | 2.50 | 2.50 | R | This work |
| 2011 01 06.0 | 3.071 | 2.244 | 11.6 | 0.94 | 1.27 | 2.55 | 2.50 | 2.50 | R | This work |
| 2011 01 09.9 | 3.065 | 2.209 | 10.7 | 0.95 | 1.29 | 2.59 | 2.49 | 2.47 | R | This work |
| 2010 12 27.9 | 3.085 | 2.338 | 13.7 | 1.26 | 2.2 | 4.4 | – | – | V | Jewitt et al. (2011) |
| 2011 01 04.9 | 3.073 | 2.254 | 11.9 | 1.00 | 1.5 | 3.0 | – | – | V | |
| 2010 12 12–2011 02 02 | 3.107–2.984 | 2.526–2.130 | 16.3–11.4 | – | – | – | 15–49 | | Rc | Ishiguro et al. (2011) |
| 2010 12 14–2010 12 15 | 3.104 | 2.496 | 16.0 | – | 200 | – | 60 | | V | Bodewits et al. (2011) |
| 2010 12 12 | 3.107 | 2.529 | 16.4 | 0.96 | 1.4 | 3.0 | – | – | V | Hsieh et al. (2012) |

(Harris and Lagerros, 2002)

$$D = \frac{1329}{\sqrt{p_v}10^{0.2\, m_a(1,1,0)}}, \quad (5)$$

where $D$ is the diameter of asteroids in kilometers. Using formulas (4) and (5), we obtained the asteroid effective radii 51.2 ± 3.0 and 50.6 ± 3.0 km, respectively. These values are close to estimations presented in the literature (Tedesco and Desert, 2002; Bauer et al., 2012; Licandro et al., 2011).

### 4.2. Dust production of the asteroid

We also used our observations of the asteroid obtained in period of "activity" (see Table 1) to estimate the dust production of the object. We used different methods for calculation of this parameter. We calculated the effective scattering cross section of the coma from equation (Jewitt et al., 2011)

$$C_C = C_n\left(10^{\Delta m} - 1\right), \quad (6)$$

where $C_n = \pi R_N^2 = 0.92 \times 10^4$ km$^2$ is the geometric cross section of the nucleus (we use aperture radius 1.6 arcsec), $\Delta m = m_n - m$, where $m_n$ is absolute R magnitude of the nucleus and $m$ is total R-band magnitude measured within a 66.72 arcsec radius aperture.

We further estimated the mass of dust particles in the coma from the scattering cross section using the relation

$$M_d = \rho \bar{a} C_C, \quad (7)$$

where $\rho$ is the particle density, taken to be $\rho = 2000$ kg m$^{-3}$, and $\bar{a} = 1$ μm is the average particle radius in the coma. We used these parameters to compare our value of radius with results obtained in article by Jewitt et al. (2011).

In addition, we calculated the dust production rate by using the methods described by Newburn et al. (1981) and Newburn and Spinrad (1985) as well as Weiler et al. (2003), taking into account the discussion by Fink and Rubin (2012). We used a density from 1000 to 2000 kg m$^{-3}$ (Moreno et al., 2011; Jewitt et al., 2011). For our estimation, we used the dust size distribution function in the form of $f(a) \sim a^{-n}$ for two variants of parameter $n$. Specifically, we considered $n=3.0$ (Moreno et al., 2011) and $n=3.5$ (Ishiguro et al., 2011). In calculation, we considered the visual geometric albedo of asteroid (Tedesco and Desert (2002)) and used the distribution with the lower and upper limits of dust grain radii taken at 0.8 μm and 5 cm (Moreno et al., 2011; Ishiguro et al., 2011), respectively. The outflow dust velocity ranged from 50 to 80 m s$^{-1}$ (Moreno et al., 2011; Ishiguro et al., 2011). Our obtained results and results from the literature are presented in Table 3.

The determination of the mass of dust escaping from Scheila to the interplanetary space by various authors differs up to almost two orders of magnitude. Our estimate, yielding the ejected mass from $2.5 \times 10^7$ to $3.4 \times 10^7$ kg, roughly agrees with the values of $4.4 \times 10^7$ and $3.0 \times 10^7$ kg found by Jewitt et al. (2011) and $3.0 \times 10^7$ kg by Hsieh et al. (2012). Both these teams however used the observations performed in V-filter, in contrast to our R-filter observations.

Ishiguro et al. (2011) as well as Bodewits et al. (2011) obtained an order of magnitude larger mass of the dust escaped from Scheila. Some discrepancy can be caused from the different parameters used in the calculations. For example, the assumed value of albedo varied from 0.038 to 0.1, densities of the particles from 1670 to 2000 kg m$^{-3}$, dust size from 0.1 μm to 10 cm, and assumed ejection velocity from 55 to 100 m s$^{-1}$.

When comparing the results, it is necessary to distinguish between the total "excavated" material and that "ejected away", into the interplanetary space. For example, Moreno et al. (2011) estimated the excavated dust mass of $2 \times 10^{10}$ kg, which is much larger than the amount observed in the object's coma and, obviously, escaping into the interplanetary space. (The latter values are given in Table 3.) Bodewits et al. (2014) estimated that only about 1–10% of the excavated material escapes into the interplanetary space (and is photometrically observed). The rest of material falls back on the surface changing its optical properties.

## Potential meteoroid streams crossed

### Cometary streams

Considering all known periodic comets from the Catalogue of Cometary Orbits (Marsden and Green, 2005), we calculated their approaches to Scheila's orbit. If there is such an approach within 0.15 AU (the maximum distance of the orbit of the cometary parent body of well-known meteor shower from the Earth's orbit; Neslušan et al., 1998) and comet produces a meteoroid stream, then we can suppose that Scheila passes through the corridor of this stream and can collide with its meteoroids.

Taking into account the result by Larson et al. (2010) that Scheila was impacted on December 3, 2010, roughly, or perhaps a short time earlier, we regard December 3.5, 2010, as the latest possible date of the impact and are interested in the events 10 days before this critical time. In this context, we found two approaches of periodic comets, with the orbital period up to 1000 years, to Scheila's orbital arc corresponding to the time interval from ten days before until the latest supposed time of collision. Specifically, comets 127P and P/2005 K3 passed within minimum distances of 0.130 and 0.034 AU, respectively, to Scheila's orbit.(Within ten days after the critical time, comet P/2003 O2 passed in the minimum distance of 0.147 AU.)

### Asteroidal streams

Although the asteroidal streams do not seem to be so numerous than cometary, some asteroids still produce their streams. This circumstance motivated us to search also for the potential asteroidal streams, the corridors of which could be crossed by Scheila. To perform this search, we used the MPCORB database[2] downloaded on June 11, 2015. It contained 686 093 orbits in total and 499 763 orbits determined from the observations covering the period of at least three oppositions and with the perihelion and aphelion of nominal orbit enabling the minimum distance to the Scheila's orbit less than 0.02 AU. We consider this limit in the case of asteroids, because the asteroidal streams are expected to be less dispersed than their cometary counterparts. The value of 0.02 AU is the minimum distance between the orbit of the Geminid parent, (3200) Phaethon, and the Earth's orbit.

We found 108 approaches, within 0.02 AU, to the Scheila's orbital arc corresponding to its position from ten days before until the supposed time of collision. If we consider a shorter orbital arc of Scheila, which corresponds to its position from two days before until the time of collision, then there were 25 such approaches within 0.02 AU. Specifically, the orbits of minor planets Nos. 37257, 63520, 67241, 82009, 83469, 91906, B5596, H1994, P9337, R7003, S0122, Y1562, Y4649, Y7882, Z2557, b3473, c7996, d1643, d3266, d6361, f1223, K01U57W, K08R77R, K08SS4W, and K14D91K approached the above-mentioned Scheila's orbital arc.

The number of 108 is a quite large. However, the asteroids (3200) Phaethon and 196 256 (2003 EH1), which are the parent bodies of two known asteroidal showers, Geminids and Quadrantids, respectively, move in the orbits with the perihelion distance lower than 0.2 AU (Phaethon) or the perihelion distance periodically changes and decreases below this limit during a certain evolutionary period (2003 EH1) (Neslušan et al., 2013). The proximity of the perihelion near the Sun causes a stress of the material in asteroidal surface layer. Breaking of the material due to the stress seems to be the mechanism leading to an occurrence of a meteoroid stream. Among the objects participating in the found 108 approaches to the Scheila's orbit, none has the perihelion distance shorter than 0.2 AU. A collision of Scheila with a meteoroid of asteroidal origin does not seem to be very probable.

### Probable boulder stream of Příbram and Neuschwanstein

Two meteorites with the known orbit, Příbram falling on April 7, 1959 (Ceplecha, 1961), and Neuschwanstein falling on April 6, 2002 (Oberst et al., 2004), moved in very similar orbits before colliding with our planet (Spurný et al., 2003). Most probably, these meteorites were the members of stream of larger boulders orbiting the Sun in a common corridor (Kornoš et al., 2008). Since the boulder stream is supposed to contain a less number of larger bodies, these do not collide with the Earth every year as the particles of a common meteoroid stream. Nevertheless, the case of Příbram and Neuschwanstein indicates that a collision is possible, whereby we can expect that the boulder stream of these two meteorites is not sole boulder stream in the interplanetary space. It is only the single known such stream at the present. We mean, the possibility of collision of Scheila with an object from a boulder stream should also be considered in our analysis.

We calculated the approach of Scheila to the orbit of Příbram (Neuschwanstein). It occurred that Scheila actually approached this orbit to the relatively small minimum distance of 0.083 AU (0.149 AU). However, the approach became in a point far from the supposed site of collision. The boulder stream of Příbram cannot account for the Scheila's outburst.

## Close approaches of small bodies

Likely, the most probable impactor on Scheila was a minor asteroidal object from the asteroid belt. And, likely, the impactor was not discovered and recorded to a database before the event. Nevertheless, to surely exclude this possibility or the possibility that it could be a satellite of a known binary asteroid, we calculate the minimum distance of known small objects from Scheila in the period from 10 days before to 10 days after the supposed time of collision on December 3.5, 2010.

No extremely close approach of any periodic-comet (up to 1000 years) to Scheila in the supposed time of collision within 0.15 AU was found. Concerning the known, three-opposition asteroids, 109 (72, 33, 10, and 3) approaches within the distance of 0.15 AU (0.125, 0.10, 0.075, and 0.05 AU, respectively) were found. The closest approach was found for minor planet W8706. However, the minimum distance between this object and Scheila was 0.022 AU, too large the W8706 or even its potential satellite could be the candidate for the impactor.

---

[2] http://www.minorplanetcenter.net/iau/MPCORB.html.

## 7. Future passages through the collisional point

If the impactor was a boulder released in a not very distant past from a periodic comet or minor planet, which produces a meteoroid stream around its orbit, or was a member of boulder stream, then Scheila can obviously be impacted by another stream member when it will cross the corridor of the stream again. Of course, the probability of another impact of a larger object is extremely small. However, there is expected a large number of tiny particles in a typical stream, which could still impact the Scheila's surface and blast into the surrounding space a small, but possibly detectable amount of dust.

Because of this possibility, we integrated Scheila's motion, considering the perturbations of all big planets, to the future and calculated the closest approaches to the supposed point of collision for three next orbital revolutions. These approaches will occur on

(1) December 6.7, 2015 (JDT = 2 457 363.2),
(2) December 9.9, 2020 (TJD = 2 459 193.4), and
(3) December 16.0, 2025 (TJD = 2 461 025.5).

Fortunately, the observational conditions will be good in all three periods around these dates, especially from the northern hemisphere of the Earth.

## Results

We found that:

(1) The morphology of outburst of the asteroid changed during the period of our observation.
(2) The radius of the asteroid is estimated to be $51.2 \pm 3.0$ and $50.6 \pm 3.0$ km.
(3) The mass of dust ejecta is estimated to be $(2.5-3.4) \times 10^7$ kg. This estimate was obtained considering the different parameters of the size, density, and velocity of dust grains.
(4) In the 10-day time interval before the latest estimated time of collision between the Scheila and an impactor, Scheila passed closely (within 0.15 AU) around the orbits of two periodic comets, 127P and P/2005 K3. If these comets produce their meteoroid streams, the impactor could be a member of one of these streams.
(5) In the 10-day time interval before the collision, Scheila also passed closely (within 0.2 AU in this case) around the orbits of 108 other main-belt asteroids. However, no perihelion distance of these objects was short and, thus, any thermal stress of surface material leading to a production of potential meteoroid stream could be expected. Most probably, the impactor did not originate from an asteroidal stream.
(6) In near future, Scheila will again pass near the collision point on December 6.7, 2015 (JDT = 2 457 363.2), December 9.9, 2020 (TJD = 2 459 193.4), and December 16.0, 2025 (TJD = 2 461 025.5).

## Acknowledgments

This work has been supported, in part, by the VEGA – the Slovak Grant Agency for Science (Grant nos. 2/0031/14 and 2/0032/14) and the implementation of the project SAIA.